\documentclass[twocolumn]{autart}    

\usepackage{amsmath}
\usepackage{amsfonts}
\usepackage{amssymb}
\usepackage{mathtools}
\usepackage{color}  
\usepackage{circuitikz}
\usepackage{pgfplots}
\usepackage{pgfplotstable}
\usepgfplotslibrary{patchplots}
\usepgfplotslibrary{groupplots}

\pgfplotsset{compat=1.5}

\newcommand\RE{\mathbb{R}}

\newcommand\setto{\rightrightarrows}

\DeclareMathOperator{\sign}{sign}

\DeclareMathOperator{\diag}{Diag}

\begin{document}

\begin{frontmatter}
\title{Dissipativity analysis of negative resistance circuits}

\thanks[footnoteinfo]{The research leading to these results has received funding from the European
Research Council under the Advanced ERC Grant Agreement Switchlet n.670645.}

\author[cam]{F\'{e}lix A. Miranda-Villatoro}\ead{fam48@cam.ac.uk},    
\author[cam]{Fulvio Forni}\ead{f.forni@eng.cam.ac.uk},               
\author[cam]{Rodolphe Sepulchre}\ead{r.sepulchre@eng.cam.ac.uk}  

\address[cam]{University of Cambridge, Department of Engineering. Trumpington Street,
Cambridge, CB2 1PZ.}  


\begin{abstract}
This paper deals with the analysis of nonlinear circuits that 
interconnect passive elements (capacitors, inductors, and resistors)
with nonlinear resistors exhibiting a range of {\it negative} resistance.
Such active elements are necessary to design circuits that switch and oscillate. We generalize 
the classical passivity theory of circuit analysis to account for such non-equilibrium behaviors.
The approach closely mimics the classical methodology of (incremental) dissipativity theory, but with 
dissipation inequalities that combine {\it signed} storage functions and {\it signed} supply rates
to account for the mixture of passive and active elements.
    \end{abstract}

\end{frontmatter}

\section{Introduction}

The concept of passivity is a foundation of circuit theory \cite{anderson2006}. It led to the generalized 
concept of dissipativity  \cite{willems1972}, \cite{willems1972b},  which has become a foundation of  
nonlinear system theory \cite{hill1980,schaft2010}. 
Yet  the applications of nonlinear system theory have been dominated by mechanical
and electro-mechanical systems \cite{brogliato2007}, \cite{Desoer2009}, \cite{ortega1998}, \cite{sepulchre1997},
with significantly less attention to nonlinear circuits \cite{brayton1964,Camlibel2002}. 

Starting with the seminal work of Chua \cite{chua1980} and the textbook of Chua and Desoer \cite{chua1987}, the
research on nonlinear circuits has somewhat diverged from the research on nonlinear dissipative
systems. The emphasis in nonlinear circuit theory has been on non-equilibrium behaviors
whereas the focus of  dissipativity theory is an interconnection framework for systems that converge
to equilibrium.

Negative resistance devices are the essence of non-equilibrium behaviors such as 
switches \cite{chen2009}, \cite{goto1960}, \cite{kennedy1991}, nonlinear
oscillations \cite{hu1986}, \cite{li2000}, or chaotic behavior \cite{kennedy1993}, \cite{saito1995}. 
In contrast, dissipativity theory is a stability theory for physical systems that only dissipate energy and that relax to equilibrium
when disconnected from an external source of energy. 

The present paper is a step towards generalizing passivity theory to the analysis of negative resistance circuits. In the spirit of passivity theory, we seek to analyze nonlinear circuits through 
dissipation inequalities that are preserved by interconnection. 

The two basic elements of dissipativity theory are the storage function and the supply function.
A dissipative system obeys a dissipation inequality, which expresses that the rate of change
of the storage does not exceed the supply. The physical interpretation is that the storage is 
a measure of the internal energy, whereas the integral of the supply is a measure of the supplied energy.
For stability analysis purposes, the storage becomes a Lyapunov function.

The approach in this paper is based on two modifications of the basic theory. First, the analysis is in terms
of {\it incremental} variables, that is, differences of voltages and currents rather than voltages and currents.
 Incremental analysis is classical in nonlinear circuit theory. Starting with the seminar work of \cite{lohmiller1998}, incremental
 analysis has  also been increasingly used in nonlinear stability theory  \cite{angeli2002}, 
\cite{forni2014b}, 
and in nonlinear dissipativity theory \cite{forni2013}, \cite{proskurnikov2015}, \cite{stan2007}, \cite{schaft2013}.
Second, we allow for
dissipation inequalities that combine {\it signed} storage functions and {\it signed} supply rates.
Signed storage functions have the interpretation of a difference of energy stored in different storage
elements whereas signed supply rates account for ports that can deliver rather than absorb energy.

For analysis purposes, the interconnection theory developed in the present paper makes contact with the 
dominance theory recently proposed in \cite{forni2017}, \cite{Forni2017b}. Signed Lyapunov functions
with a restricted number of negative terms are used to prove convergence to low-dimensional
dynamics that dominate the asymptotic behavior. A one-dimensional dominant behavior is sufficient
to model bistable switches whereas a two-dimensional dominant behavior is sufficient to model
nonlinear oscillators. Combined with the interconnection theory of this paper, dominance theory
opens the way to analysis of nonlinear switches and nonlinear oscillators in large nonlinear circuits.

We deliberately restrict the scope of the present paper to nonlinear circuits with negative resistance
to facilitate a concrete interpretation of the results. Not surprisingly,
the concepts are not restricted to electrical circuits  and have a more general interpretation
in the general framework of dissipativity theory. For concreteness,
the entire paper is restricted to the passivity supply, an inner product  between currents and voltages,
with the convenient interpretation of electrical power.

The paper is organized as follows. Section \ref{section:motivation} deals with the dissipation properties
of negative resistance devices and Section \ref{section:differential} extends dominance theory 
in an incremental framework that is suitable for the analysis of circuits with piecewise linear characteristics. 
In Section \ref{section:circuits:lossless:lure} we analyze basic electrical
switches and oscillators with one or two storage elements, whereas Section \ref{section:circuits:connection} covers the 
design of coupling networks that allows us to interconnect circuits with different signatures in the supply rates.

{\small \textbf{Preamble.}

The circuits studied in this paper are built from interconnections of \emph{linear passive} 
elements, such as capacitors and inductors, and \emph{nonlinear active} resistors. In concrete,
the time evolution of the family of circuits studied here is described by the state-space model
\begin{equation}
  \Sigma: \begin{cases}
    \dot{x} = f(x) + B u  \quad x(0) = x_{0}
  \\
  y = C x + D u
  \end{cases}
    \label{eq:circuit:ss}
\end{equation}
where $x \in \RE^{n}$ is the state of the system and $u, y \in \RE^{m}$
are the so-called manifest variables. 
For electrical circuits, the manifest variables are conjugated in terms of voltages 
$v$, and currents $i$,
that is, the inner product $u^{\top} y$ has units of power.
The map $f: \RE^{n} \to \RE^{n}$ is Lipschitz continuous and models interactions between linear storage elements 
and nonlinear resistors.
Moreover, the matrices $B$, $C$, and $D$ are of the appropriate dimensions and such that 
the system is well-posed.
Henceforth, every circuit in this paper is assumed to be of the form \eqref{eq:circuit:ss}. 
In what follows we will adopt a \emph{differential} (or incremental) approach, that is, 
we will study circuit properties by looking at the difference between trajectories. For simplicity, we 
denote the difference between any two generic signals $w_{1}, w_{2}$ as
$\Delta {w} := w_{1} - w_{2}$. In this way, 
the mismatches between any two states/currents/voltages are denoted as 
$\Delta {x}$, $\Delta {i}$ and $\Delta {v}$ respectively.
Finally, we will use symmetric matrices $P \in \RE^{n \times n}$ constrained
to have inertia $(p, 0, n-p)$, that is, with $p$ negative eigenvalues and $n-p$ positive 
eigenvalues.}

\section{Signed supply rates for nonlinear resistors}
\label{section:motivation}

\begin{figure}[htpb]
  \centering
  \includegraphics{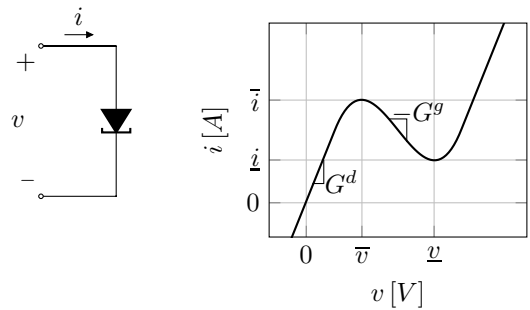} \quad
  \raisebox{-8.0ex}{\begin{tikzpicture}
\begin{axis}[
    xtick = {-1.78, -0.7, 0.7},
    xticklabels = {0, $\overline{v}$, $\underline{v}$},
    ytick = {-1.7, -0.7, 0.7},
    yticklabels = {0, $\underline{i}$, $\overline{i}$},
x grid style={},
xlabel={$v \, [V]$},
xmajorgrids,
xmin=-2.5, xmax=2.5,
ylabel={$i \, [A]$},
ymajorgrids,
ymin=-2.5, ymax=2.5,
scale=0.5
]
\addplot [thick, black, forget plot] table [x=Y1, y=Y2, col sep=comma]{vi_nr_smooth.csv};
\addplot [black, no markers] coordinates {
(-1.44, -0.7) 
(-1.44, -1.25) 
(-1.66, -1.25)
};
\addplot [black, no markers] coordinates {
(-0.17, 0.26) 
(0.17, 0.26) 
(0.17, -0.28)
};
\node[] at (axis cs: -1.1, -1.2, 0.16) {$G^{d}$};
\node[] at (axis cs: 0.4, 0.4, 0.16) {$-G^{g}$};
\end{axis}
\end{tikzpicture}
}
  \caption{Slope-bounded voltage-current characteristic of a tunnel diode. Tunnel diodes are 
    (incrementally) negative resistance devices. The region of 
  negative slope is called the \emph{active} region.}
  \label{fig:tunnelDiode}
\end{figure}

The nonlinear element shown in Figure \ref{fig:tunnelDiode} is a fundamental element
of nonlinear circuits. The voltage range where the nonlinear characteristic has a negative slope models
an element that can deliver energy rather than dissipating energy. Such an element is called
{\it active} in contrast to {\it passive} elements that can only absorb energy.
We follow the common terminology of {\it negative resistance} device \cite{chua1983}, \cite{kaplan1968}, with the usual caveat
that {\it negative} refers to the {\it increment }$\Delta {v}$ rather than to the value of the voltage $v$. 
A more precise (but also heavier) terminology would be {\it negative incremental (or differential)}
resistance. The analysis in this paper will be exclusively in terms of {\it incremental} quantities,
which is common practice in nonlinear circuit theory.

We are motivated by the property that this nonlinear element satisfies the two inequalities
\begin{subequations}
 \begin{align}
   0 & \leq \Delta {i} \Delta {v} + G^{g} (\Delta {v})^{2}  
  \label{eq:tunnel:supply:1}
  \\
  0 & \leq -\Delta {i} \Delta {v} + G^{d} (\Delta {v})^{2} 
  \label{eq:tunnel:supply:2}
\end{align}
  \label{eq:tunnel:supply}
\end{subequations}
where $G^{d} > 0$ and $-G^{g} < 0$ represent, respectively, the maximum positive slope and negative slope of the
voltage-current characteristic of Figure \ref{fig:tunnelDiode}. Both inequalities have an obvious energetic interpretation: the first inequality
expresses the shortage of passivity of the element: the element becomes passive when connected in parallel with a resistor
of resistance lesser than ${1/G^g}$. The second inequality expresses the shortage of anti-passivity of the element: the element
becomes purely a source of energy when connected to a negative resistance larger than $-1/G^d$.  

In the language of dissipativity theory \cite{willems1972}, both inequalities are dissipation inequalities of the form $ \sigma(\Delta {i}, \Delta {v}) \ge 0$
for the family of quadratic supply rates
  \begin{equation}
  \label{eq:supply:1}
  \sigma(\Delta {i}, \Delta {v}) = 
  \begin{bmatrix}
    \Delta {i} \\ \Delta {v}
  \end{bmatrix}^{\top}
  \begin{bmatrix}
    \mathcal{Q} & \mathcal{I}
    \\
    \mathcal{I} & \mathcal{R}
  \end{bmatrix}
  \begin{bmatrix}
    \Delta {i} \\ \Delta {v}
  \end{bmatrix}
\end{equation}
where the signature matrix $\mathcal{I} \in \RE^{m \times m}$ is a diagonal matrix
with $\pm 1$ in the main diagonal 
$\mathcal{I} = \diag [ \pm 1, \pm 1, \dots, \pm 1 ]$,
and $\mathcal{Q} \in \RE^{m \times m}$,
$\mathcal{R} \in \RE^{m \times m}$ are symmetric matrices. In the special case $\mathcal{I}=I$, this family 
of supply rates characterize incrementally passive elements with an excess or a shortage
of passivity  in the external variables \cite{sepulchre1997}. When $\mathcal{Q} = 0$, 
the dissipativity property $\sigma(\Delta {i}, \Delta {v}) \ge 0$ is also equivalent
to the monotonicity of the voltage-current characteristic $i = g(v)$ \cite{bauschke2011}. 
The map $g$  is called
strongly monotone for $\mathcal{R} >0$, hypomonotone for $\mathcal{R} < 0$ and monotone 
for $\mathcal{R} = 0$.

We call (\ref{eq:supply:1}) a {\it signed} passivity supply
rate to stress that the only difference with respect to the conventional passivity supply is the signature
matrix  $\mathcal{I}$ generalizing the conventional identity matrix $I$.

\begin{figure}[htpb]
  \centering
  \includegraphics{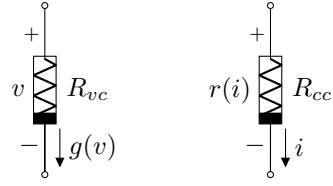}
  \caption{Voltage-controlled resistor (left) and current-controlled resistor (right). The functions
  $g$ and $r$ are assumed singled-valued and Lipschitz continuous. If $g$ or $r$ are monotone increasing 
  then the resistor is passive, otherwise it is active.}
  \label{fig:nr:vc:cc}
\end{figure}

The element  in Figure \ref{fig:tunnelDiode} is called a  voltage-controlled resistor,
Figure \ref{fig:nr:vc:cc} (left). Namely, the current flowing through a voltage-controlled resistor
is a singled-valued function of the voltage across its terminals: $i = g(v)$. The nonlinear resistor is passive
when the function $g: \RE \to \RE$ is monotone increasing,  otherwise it is active. 
It follows from \eqref{eq:tunnel:supply} that  whenever $G^{d} \neq G^{g}$, a voltage-controlled resistor  fulfills
\begin{equation}
  \label{eq:supply:nr}
  0 \leq  
  \begin{bmatrix}
    \Delta {i} \\ \Delta {v}
  \end{bmatrix}^{\top}
  \begin{bmatrix}
    \mathcal{Q} & \mathcal{I}
    \\
    \mathcal{I} & \mathcal{R}
  \end{bmatrix}
  \begin{bmatrix}
    \Delta {i} \\ \Delta {v}
  \end{bmatrix}
\end{equation}
where $\mathcal{I} = \sign(G^{d} - G^{g})$, $\mathcal{Q} = -\frac{2}{\vert G^{d} - G^{g} \vert}$
and $\mathcal{R} = \frac{2 G^{g} G^{d}}{\vert G^{d} - G^{g} \vert}$. 

The dual element is the current-controlled resistor defined by 
a singled-valued function of its flowing current: $v = r(i)$. 
An active current-controlled resistor satisfies the sector condition
\begin{equation}
  \label{eq:sector:ccnr}
  -R^{g} (\Delta {i})^{2} \leq \Delta {i} \Delta {v} \leq  R^{d} (\Delta {i})^{2}
\end{equation}
Equivalently, a current-controlled resistor satisfies \eqref{eq:supply:nr} with 
$\mathcal{I} = \sign(R^{d} - R^{g})$, $\mathcal{Q} = \frac{2 R^{g} R^{d}}{ \vert R^{d} - R^{g} \vert}$
and $\mathcal{R} = - \frac{2}{ \vert R^{d} - R^{g} \vert}$.
Both types of controlled resistors appear naturally in devices such as tunnel diodes, DIAC's or neon lamps.
Additionally, they can be built from off-the-shelf components like transistors and operational amplifiers
\cite{chua1983}, \cite{kaplan1968}.

\begin{figure}[htpb]
  \centering
  \includegraphics[trim={0.3cm, 0.35cm, 0.35cm, 0.35cm}, clip]{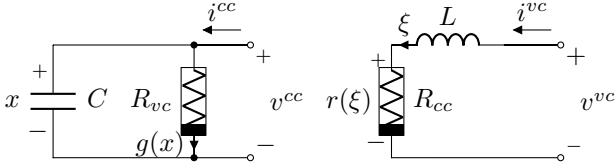}
  \caption{Basic prototype circuits of a current-driven (left) and a voltage-driven (right) 
  $1$-passive circuit. The resistors $R_{vc}$ and $R_{cc}$ are voltage-controlled and current-controlled
resistors respectively.}
  \label{fig:basic:sw:vc}
\end{figure}

Describing negative resistors in terms of dissipation inequalities opens the way to the
use of dissipativity theory to characterize circuit interconnections.
As an illustration, consider the parallel interconnection
of a voltage-controlled negative resistance element with a capacitor (Figure \ref{fig:basic:sw:vc}, left). 
Let $i^{c}, v^{c}$ and $i^{r}, v^{r}$ be the currents and voltages associated to the capacitor and the 
controlled resistor, respectively.
The capacitor is a classical lossless element that satisfies the power-preserving equality
  \begin{equation}
  \label{eq:supply:capacitor}
\frac{d}{dt} {C \frac {(\Delta v^{c})^2}{2}} = \Delta {v^{c}} \Delta {i^{c}}
  \end{equation}
In the language of dissipativity theory,   the quantity on the left-hand side is the time-derivative of the 
{\it storage} $C \frac{(\Delta v^{c})^2}{2}$.
The negative resistance element satisfies $-\Delta {v^{r}} \Delta {i^{r}} + G^{d} (\Delta v^{r})^{2}  \ge 0$.
The parallel interconnection defined by $v^{cc}=v^{c}=v^{r}$ and $i^{cc}=i^{c}+i^{r}$~\footnote{The superindices in the variables 
  $i^{cc}$ and $v^{cc}$ indicate that the port under consideration is current-driven. In a similar way, $i^{vc}$ and
$v^{vc}$ will denote the variables associated to a voltage-driven port.} satisfies  the dissipation (in)equality
  \begin{equation} 
    -\frac{d}{dt} {C \frac {(\Delta {v}^{cc})^{2}}{2}}  \le  -\Delta v^{cc} \Delta i^{cc} +  G^{d} (\Delta {v}^{cc})^{2}
    \label{eq:sw1:ineq:dissipation}
  \end{equation}
The quantity that appears on the left hand-side is the time-derivative of a {\it negative} storage. 
More generally, the storage functions in this paper will be quadratic forms defined by a symmetric
matrix $P=P^T$ with $p$ negative eigenvalues (and $n-p$ positive eigenvalues). Such {\it signed} storage
functions generalize the conventional {\it positive definite} storages of passivity theory. Positive definite
storages are natural candidates for the stability analysis of closed equilibrium systems. 
In its incremental form, stability analysis appears in the literature under different names, 
including  {\it contraction} theory \cite{lohmiller1998}, {\it incremental} stability analysis \cite{angeli2002}, 
or differential Lyapunov analysis \cite{forni2014b}.
{\it Signed} storages generalize this stability analysis for non-equilibrium behaviors   characterized by a low-dimensional  asymptotic behavior.
This generalization is the topic of dominance analysis, reviewed in the next section.  

\section{Differential dissipativity}
\label{section:differential}

\subsection{Dominant systems}
\label{section:dominance}

Dominance theory extends stability analysis to non-equilibrium behaviors. The approach
is based on the intuitive idea that the long run behavior of the system
is dictated by low-dimensional dynamics, identified through the study of the system
linearization \cite{forni2014b}, \cite{forni2017}, \cite{Forni2017b}.
In what follows we adapt the differential approach of \cite{Forni2017b} into 
an incremental setting.

\begin{defn}
  Let $f: \RE^{n} \setto \RE^{n}$ be a Lipschitz continuous map. A system of the form 
  \begin{equation}
    \dot{x} \in f(x), \quad x \in \RE^{n},
    \label{eq:inclusion}
  \end{equation}
  is $p$-dominant with rate $\lambda \geq 0$ if there exists a matrix
  $P = P^{\top} \in \RE^{n \times n}$ with inertia $(p, 0, n-p)$ such that 
  \begin{equation}
    \begin{bmatrix}
      \Delta {\dot{x}} \\ \Delta {x}
    \end{bmatrix}^{\top}
    \begin{bmatrix}
      0 & P 
      \\
      P & 2 \lambda P + \varepsilon I
    \end{bmatrix}
    \begin{bmatrix}
      \Delta {\dot{x}} \\ \Delta {x}
    \end{bmatrix}
    \leq 0.
    \label{eq:dominance}
  \end{equation}
  The property is strict if $\varepsilon > 0$.
  \label{def:dominance}
\end{defn}

When $P$ is positive definite, \eqref{eq:dominance} 
becomes the incremental analogue of the classical Lyapunov inequality, meaning that any two trajectories
converge to each other with decay rate at least $\lambda \geq 0$, \cite{Boyd1994}. 
When $f$ is a differentiable map, \eqref{eq:dominance} 
reduces to the simple matrix inequality
\begin{equation}
  \frac{\partial f(x)}{\partial x}^{\top} P + P \frac{\partial f(x)}{\partial x}
  + 2 \lambda P \leq -\varepsilon I,
  \label{eq:dominance:smooth}
\end{equation}
which provides a basic test for dominance, \cite{forni2017}, \cite{Forni2017b}.

\begin{thm}
  Let $f: \RE^{n} \to \RE^{n}$ be a differentiable map. The closed system
  \eqref{eq:inclusion} is $p$-dominant if and only if, there exists a matrix $P = 
  P^{\top}$
  with inertia $(p, 0, n-p)$ such that \eqref{eq:dominance:smooth} holds.
  \label{thm:dominance:smooth}
\end{thm}
\begin{pf}
  First assume that \eqref{eq:inclusion} is $p$-dominant. Expanding the left-hand side of
  \eqref{eq:dominance} and dividing by $\Vert \Delta {x} \Vert^{2} \neq 0$ yields,
  \begin{displaymath}
    \frac{\Delta {f}^{\top} P \Delta {x} +
      \Delta {x}^{\top} P \Delta {f}
      + 2 \lambda \Delta {x}^{\top} P \Delta {x} + 
    \varepsilon \Delta {x}^{\top} \Delta {x}}{\Vert \Delta {x} \Vert^{2}} \leq 0.
  \end{displaymath}
  By letting $\delta_{x} = \lim_{\Delta {x} \to 0} \frac{\Delta {x}}{\Vert \Delta {x} \Vert}$ we
  arrive to \eqref{eq:dominance:smooth}.
  For the converse statement, let $x (\alpha) = \alpha x_{1} + (1 - \alpha) x_{2}$ and 
  let $\phi:\RE \to \RE$ be such that
  \begin{multline*}
    \phi(\alpha) = 2 \left( f(x(\alpha)) - f(x_{2}) + \lambda ( 
    x(\alpha) - x_{2}) \right)^{\top} P \Delta {x} 
    \\
    + \varepsilon (x(\alpha) - x_{2})^{\top} 
    \Delta {x}
  \end{multline*}
  where $\Delta x = x_{1} - x_{2}$. Hence,
  \begin{multline*}
    \frac{d \phi(\alpha)}{d \alpha} = \Delta {x}^{\top} \left( \frac{\partial f(x)}{\partial x}^{\top} 
    P + P \frac{\partial f(x)}{\partial x} \right.
    \\
    \left. \phantom{\frac{1}{2}} + 2 \lambda 
  P + \varepsilon I \right) \Delta {x}
    \leq 0.
  \end{multline*}
  The above inequality implies that $\phi$ is a non-increasing function. Therefore, 
  $\phi(1) \leq \phi(0) = 0$ and \eqref{eq:dominance} follows. This concludes the proof. $\hfill\qed$
\end{pf}
The property of dominance strongly constrains the asymptotic behavior
of the system as described for the following theorem.
\begin{thm}[{\cite[Theorem 2]{Forni2017b}}]
  \label{theorem:dominance:constrain}
  Let \eqref{eq:inclusion}
  be strictly $p$-dominant with rate $\lambda \geq 0$.
  For any given $x \in \mathbb{R}^n$,  let $\Omega(x)$ be the $\omega$-limit set of $x$.
  Then the flow of \eqref{eq:inclusion} on $\Omega(x)$ is topologically equivalent to the flow of
  a $p$-dimensional system.
  \label{eq:dominance:constrain}
\end{thm}
Additionally, the following corollary becomes useful in characterizing
the asymptotic behavior of a dominant system.
\begin{cor}
  Under the assumptions of Theorem \ref{theorem:dominance:constrain}, 
  every bounded trajectory of \eqref{eq:inclusion} converges to
  \begin{itemize}
    \item A unique equilibrium point if $p = 0$.
    \item An equilibrium point if $p = 1$.
    \item A simple attractor if $p = 2$.
  \end{itemize}
  \label{corollary:behavior}
\end{cor}
Summing up, closed dynamic systems with smaller degrees of dominance will show simpler behaviors
compared with systems with higher degrees. The following subsection extends the property of
dominance to open systems under the framework of dissipative systems.

\subsection{Signed dissipation inequalities}

Dissipativity theory \cite{willems1972}, \cite{willems1972b} is grounded in dissipation inequalities, 
which generalize the physical characterization of a passive circuit as a system that can only absorb energy:
the variation of energy {\it stored} in the elements of the circuit (capacitors and inductors) is upper bounded
by the electrical power {\it supplied} to the circuit. For a linear circuit, the storage is a quadratic function of the state,
and the dissipation inequality takes the standard form
$$ \frac{d}{dt} x^{\top}Px \le - \lambda x^{\top} P x + v^{\top} i + i^{\top} v $$
The scalar $\lambda \ge 0$ determines a dissipation rate. Each pair of voltage $v_k$ and current $i_k$ appearing in
the voltage vector $v$ and voltage current $i$ determines a port of the circuit.

In matrix form, the quadratic dissipation inequality characterizing passivity reads
  \begin{equation}
    \begin{bmatrix}
      \dot{x} \\
        x
    \end{bmatrix}^{\top}
    \begin{bmatrix}
      0 & P 
      \\
      P & 2 \lambda P 
    \end{bmatrix}
    \begin{bmatrix}
      \dot{x}
      \\
      x
    \end{bmatrix}
    \leq
    \begin{bmatrix}
      v \\ i
    \end{bmatrix}^{\top}
    \begin{bmatrix}
      0 & I
      \\
    I & 0
  \end{bmatrix}
    \begin{bmatrix}
      v \\ i
    \end{bmatrix}
    \label{eq:passivity:inequality}
  \end{equation}
  
  An incremental dissipation inequality is in term of the increments rather than absolute variables:
  \begin{equation}
    \begin{bmatrix}
      \Delta \dot{x} \\
        \Delta x
    \end{bmatrix}^{\top}
    \begin{bmatrix}
      0 & P 
      \\
      P & 2 \lambda P 
    \end{bmatrix}
    \begin{bmatrix}
      \Delta \dot{x}
      \\
     \Delta  x
    \end{bmatrix}
    \leq
    \begin{bmatrix}
      \Delta v \\  \Delta i
    \end{bmatrix}^{\top}
    \begin{bmatrix}
      0 & I
      \\
    I & 0
  \end{bmatrix}
    \begin{bmatrix}
      \Delta v \\ \Delta i
    \end{bmatrix}
    \label{eq:inc-passivity:inequality}
  \end{equation}

Motivated by the signed supply rates and signed storages introduced in Section \ref{section:motivation}, we generalize  the incremental passivity dissipation inequality (\ref{eq:inc-passivity:inequality})  to  {\it signed} dissipation inequalities of the form
  \begin{equation}
    \begin{bmatrix}
      \Delta {\dot{x}} \\
      \Delta {x}
    \end{bmatrix}^{\top}
    \begin{bmatrix}
      0 & P 
      \\
      P & 2 \lambda P + \varepsilon I
    \end{bmatrix}
    \begin{bmatrix}
      \Delta {\dot{x}}
      \\
      \Delta {x}
    \end{bmatrix}
    \leq
    \begin{bmatrix}
      \Delta {v} \\ \Delta {i}
    \end{bmatrix}^{\top}
    \begin{bmatrix}
      \mathcal{Q} & \mathcal{I}
      \\
      \mathcal{I} & \mathcal{R}
  \end{bmatrix}
    \begin{bmatrix}
      \Delta {v} \\ \Delta {i}
    \end{bmatrix}
    \label{eq:dissipation:inequality}
  \end{equation}
  for an arbitrary circuit with state $x \in \RE^n$ and $m$ ports defining the current  $i \in \RE^m$ and voltage  $v \in \RE^m$.
  We only consider circuits composed of linear capacitors, linear inductors, and nonlinear resistors. The {\it signed} quadratic storage is
  determined by the symmetric matrix $P$ with $p$ negative eigenvalues and $n-p$ positive eigenvalues.  The {\it signed} supply
  is determined by the signature matrix $\mathcal{I}$. The scalar  $\lambda \geq 0$ is the dissipation rate.
  The matrices $\mathcal{Q}, \mathcal{R}$ are symmetric as in \eqref{eq:supply:1}. 
  
  \begin{defn}
    A nonlinear circuit is called {\it signed} passive if the inequality (\ref{eq:dissipation:inequality}) holds along any pair of trajectories.
    The property is strict if $\varepsilon > 0$. 
    \label{def:signed:passive}
  \end{defn}
  Definition \ref{def:signed:passive} is very close to the classical definition of incremental passivity. The only 
   difference is that (i) we consider  {\it signed} storages, i.e.  {\it differences} 
   of positive storages and (ii) {\it signed} supply rates, i.e. {\it differences} of the classical {\it passivity} supply rates.
   As illustrated in Section \ref{section:motivation}, such storages and supply rates appear naturally when considering
   circuits with both passive and active elements and ports that can both absorb and deliver energy. 
 
\subsection{Dissipative interconnections}

The central property of passivity theory is that passivity is preserved by interconnection. More precisely, port interconnections
of passive circuits are passive. In order to generalize this property to signed-passivity, we introduce the following definition.

\begin{defn}
  \label{def:dissipative:connection}
  Let $\Sigma_{a}$ and $\Sigma_{b}$ be signed-passive with a common rate $\lambda \ge 0$.
  Their interconnection is called \emph{dissipative} if
  \begin{equation}
    \label{eq:dissipative:connection}
    \Delta {i^{a}}^{\top} \mathcal{I}_{a} \Delta {v^{a}} + \Delta {i^{b}}^{\top} \mathcal{I}_{b} \Delta {v^{b}} 
    \leq \Delta {i} \mathcal{I} \Delta {v}
  \end{equation} 
  If equality holds in (\ref{eq:dissipative:connection}), then the interconnection is called {\it neutral}.
\end{defn}

The conventional passivity supply assumes $\mathcal{I}=I$. In this case, an interconnection is {\it neutral}
if 
$$  \Delta {i^{a}}^{\top} \Delta {v^{a}} + \Delta {i^{b}}^{\top}  \Delta {v^{b}} 
    = \Delta {i}^{\top} \Delta {v}
    $$
Hence, port interconnections of passive circuits are neutral. More generally, 
let us consider the port interconnection of two signed-passive systems as
\begin{align}
  \nonumber
  i^{a} & = - i^{b} + i^{cc} & i^{b} & = - i^{vc}
  \\
  v^{a} & = v^{b} + v^{vc} & v^{a} & = v^{cc}
  \label{eq:simple:pattern}
\end{align}
where we have set $i = [i^{cc \top}, i^{vc \top} ]^{\top}$ and
$v = [v^{cc \top}, v^{vc \top}]^{\top}$. Here the pairs $(i^{cc}, v^{cc})$ and $(i^{vc}, v^{vc})$ 
are associated to current-controlled and voltage-controlled ports, respectively, 
see Figures \ref{fig:basic:sw:vc} and \ref{fig:basic:osc}.
Substitution of \eqref{eq:simple:pattern} on the left-hand side of \eqref{eq:dissipative:connection}
shows that port interconnections of signed-passive systems with supplies 
sharing the same signature (i.e., $\mathcal{I}_{a} = \mathcal{I}_{b}$) 
are neutral.
Note that a circuit is closed or terminated whenever $i^{cc} = 0$ and $v^{vc} = 0$.

The question of how to realize a neutral or dissipative interconnection when interconnecting signed-passive circuits 
 is deferred    to Section \ref{section:circuits:connection}.  But the definition allows for the following generalization 
 of the passivity theorem.
 
\begin{thm}
  \label{thm:interconnection}
The dissipative interconnection of two signed-passive systems with a common dissipation rate
is signed-passive with the same rate. The storage of the interconnected system is the sum of the storages. 
\end{thm}

\begin{pf}
  Let us consider the aggregated state $x = [x_{a}^{\top}, x_{b}^{\top}]^{\top}$, and the block-diagonal
  matrix $P = \diag [P_{a}, P_{b} ]$. The sum of storages satisfies,
  \begin{multline}
    \begin{bmatrix}
      \Delta \dot{x} \\ \Delta x
    \end{bmatrix}^{\top}
    \begin{bmatrix}
      0 & P
      \\
      P & 2 \lambda P + \varepsilon I
    \end{bmatrix}
    \begin{bmatrix}
      \Delta \dot{x} \\ \Delta x
    \end{bmatrix} \leq 
    \\
    \sum_{k \in {a, b}} 
    \begin{bmatrix}
      \Delta i^{k} \\ \Delta v^{k}
    \end{bmatrix}^{\top}
    \begin{bmatrix}
      \mathcal{Q}_{k} & \mathcal{I}_{k}
      \\
      \mathcal{I}_{k} & \mathcal{R}_{k}
    \end{bmatrix}
    \begin{bmatrix}
      \Delta i^{k} \\ \Delta v^{k}
    \end{bmatrix}
    \label{eq:sum:supplies}
  \end{multline}

  Simple, yet cumbersome, computations show that the substitution of the interconnection 
  pattern \eqref{eq:simple:pattern} into \eqref{eq:sum:supplies} together with 
  the dissipativity of the interconnection yield,
  \begin{multline}
    \begin{bmatrix}
      \Delta \dot{x} \\ \Delta x
    \end{bmatrix}^{\top}
    \begin{bmatrix}
      0 & P
      \\
      P & 2 \lambda P + \varepsilon I
    \end{bmatrix}
    \begin{bmatrix}
      \Delta \dot{x} \\ \Delta x
    \end{bmatrix} \leq 
    \\
    \begin{bmatrix}
      \Delta i^{cc} \\ \Delta i^{vc} \\ \Delta v^{cc} \\ \Delta v^{vc}
    \end{bmatrix}^{\top}
    \begin{bmatrix}
      \hat{\mathcal{Q}} & \hat{\mathcal{I}}
      \\
      \hat{\mathcal{I}} & \hat{\mathcal{R}}
    \end{bmatrix}
        \begin{bmatrix}
      \Delta i^{cc} \\ \Delta i^{vc} \\ \Delta v^{cc} \\ \Delta v^{vc}
    \end{bmatrix}
    \label{eq:supply:connection}
  \end{multline}
  where $\hat{\mathcal{I}} = \diag [\mathcal{I}_{a}, \mathcal{I}_{b}]$ and
  \begin{align*}
    \hat{\mathcal{Q}} & = 
    \begin{bmatrix}
      \mathcal{Q}_{a} & - \mathcal{Q}_{a}
      \\
      - \mathcal{Q}_{a} & \mathcal{Q}_{a} + \mathcal{Q}_{b} 
    \end{bmatrix}
    &
    \hat{\mathcal{R}} & =
    \begin{bmatrix}
      \mathcal{R}_{a} + \mathcal{R}_{b} & - \mathcal{R}_{b}
      \\
      - \mathcal{R}_{b} & \mathcal{R}_{b}
    \end{bmatrix}
  \end{align*}
  and the result follows. $\hfill\qed$
\end{pf}

A key consequence of the passivity theorem is the property that when a passive system is terminated, it leads 
to a stable equilibrium system. The storage becomes a Lyapunov function for the closed system. 
The generalization of that result is as follows.

\begin{thm}
  \label{thm:dominance:closedLoop}
  Let  $\Sigma_{a}$ be a strictly signed-passive circuit with rate $\lambda > 0$ and dominance degree $p$.  
  The terminated circuit built from the dissipative interconnection of $\Sigma_{a}$  with a resistor ($\Sigma_{b}$) defines a  
  $p$-dominant system with the same rate $\lambda > 0$ provided that
  $\mathcal{Q}_{a} + \mathcal{Q}_{b} \leq 0$ and $\mathcal{R}_{a} + \mathcal{R}_{b} \leq 0$.
  \end{thm}

  \begin{pf}
    Recall that a resistor (linear or nonlinear) satisfies \eqref{eq:supply:nr}. Thus, from Theorem  \ref{thm:interconnection},
    the interconnection satisfies \eqref{eq:supply:connection}. In addition, the termination of the ports, 
    i.e., $i^{cc} = 0$ and $v^{vc} = 0$, transforms \eqref{eq:supply:connection} into
    \begin{multline*}
     \begin{bmatrix}
      \Delta \dot{x} \\ \Delta x
    \end{bmatrix}^{\top}
    \begin{bmatrix}
      0 & P
      \\
      P & 2 \lambda P + \varepsilon I
    \end{bmatrix}
    \begin{bmatrix}
      \Delta \dot{x} \\ \Delta x
    \end{bmatrix} \leq  
    \\
    \begin{bmatrix}
      \Delta i^{vc} \\ \Delta v^{cc}
    \end{bmatrix}^{\top}
    \begin{bmatrix}
      \mathcal{Q}_{a} + \mathcal{Q}_{b} & 0
      \\
      0 & \mathcal{R}_{a} + \mathcal{R}_{b}
    \end{bmatrix}
    \begin{bmatrix}
      \Delta i^{vc} \\ \Delta v^{cc}
    \end{bmatrix} \leq 0
    \end{multline*}
    and the conclusion follows directly 
    from Definition \ref{def:dominance}. $\hfill\qed$
  \end{pf}

\section{Elementary switching and oscillating circuits}
\label{section:circuits:lossless:lure}

In this section we review classical elementary circuits and illustrate their signed passivity properties.

\subsection{Switching circuits}

We start with  the parallel nonlinear $RC$ circuit  and the 
series nonlinear $RL$ circuit shown in Figure \ref{fig:basic:sw:vc}. 
For the nonlinear $RC$ circuit, we rewrite the dissipation inequality \eqref{eq:sw1:ineq:dissipation} in the matrix form with state $x = v^{c}$
\begin{multline}
  \label{eq:cc_sw:ineq}
\begin{bmatrix}
    \Delta {\dot x} \\ \Delta {x}
  \end{bmatrix}^{\top}
  \begin{bmatrix}
    0 & -\frac{C}{2}
    \\
    -\frac{C}{2} & - \lambda C
  \end{bmatrix}
  \begin{bmatrix}
    \Delta {\dot x} \\ \Delta {x}
  \end{bmatrix}
 \leq 
 \\
 \frac{1}{2}
  \begin{bmatrix}
    \Delta {i}^{cc} \\ \Delta {v}^{cc}
  \end{bmatrix}^{\top}
  \begin{bmatrix}
    0 & -1
    \\
    -1 & 2(G^{d} - \lambda C)
  \end{bmatrix}
  \begin{bmatrix}
    \Delta {i}^{cc} \\ \Delta {v}^{cc}
  \end{bmatrix}
\end{multline}
The dissipation inequality
involves the standard storage of a capacitor and the standard supply of a one port circuit,
but both with a negative signature.
 
The circuit is the port interconnection of a capacitor with a negative resistor. The interconnection 
is neutral as a port interconnection of elements with negative signature ${\mathcal I}=-1$. 
Terminating the circuit, that is, setting $i^{cc} = 0$, results in a $1$-dominant system when $G^{d} - \lambda C < 0$. 
This closed circuit has one or three equilibria. With three equilibria, one of which unstable, the circuit is an elementary
example of bistable switch.

The dissipativity analysis of the  series $RL$ circuit in Figure \ref{fig:basic:sw:vc} is  similar.
Taking as state variable $\xi$, the circuit satisfies the dissipation inequality
\begin{multline}
  \label{eq:vc_sw:ineq}
\begin{bmatrix}
    \Delta {\dot \xi} \\ \Delta {\xi}
  \end{bmatrix}^{\top}
  \begin{bmatrix}
    0 & -\frac{L}{2}
    \\
    -\frac{L}{2} & - \lambda L
  \end{bmatrix}
  \begin{bmatrix}
    \Delta {\dot \xi} \\ \Delta {\xi}
  \end{bmatrix}
 \leq 
 \\
 \frac{1}{2}
  \begin{bmatrix}
    \Delta {i}^{vc} \\ \Delta {v}^{vc}
  \end{bmatrix}^{\top}
  \begin{bmatrix}
    2(R^{d} - \lambda L) & -1
    \\
    -1 & 0
  \end{bmatrix}
  \begin{bmatrix}
    \Delta {i}^{vc} \\ \Delta {v}^{vc}
  \end{bmatrix}
\end{multline}
The circuit is a bistable switch when
$R^{d} - \lambda L < 0$. Both circuits can be seen as abstract realizations of the classical Schmitt trigger circuit in which the negative
resistor is usually made by using an operational amplifier in positive feedback \cite{miranda2018a}.

\subsection{Oscillating circuits}

\begin{figure}[hptb]
\begin{center}
  \includegraphics[trim={0.5cm, 0.3cm, 0.45cm, 0.3cm}, clip]{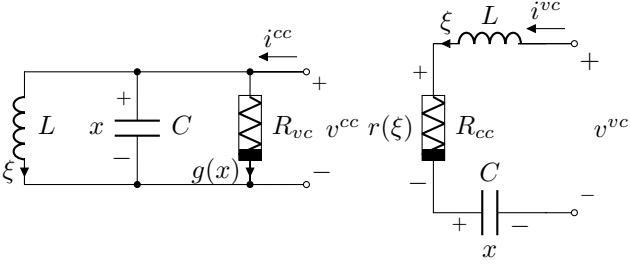}
  \end{center} \vspace{-3mm}
  \caption{Basic prototype circuits of a current-controlled (left) and a voltage-controlled (right) 
   signed-passive circuits with degree of dominance $2$. } 
  \label{fig:basic:osc}
\end{figure}

We proceed with the analysis of the nonlinear RLC circuits 
shown in Figure \ref{fig:basic:osc}.

The parallel  nonlinear $RLC$ circuit is the port interconnection of the nonlinear $RC$
circuit in the previous section with a lossless inductor. The port interconnection is neutral
as an interconnection of two circuits with supply signature ${\mathcal I} = -1$. The total storage
is the sum of two negative storages 
$$ - \frac{C}{2} (\Delta x)^2 - \frac{L}{2} (\Delta \xi)^2. $$
Defining the state $\Delta z = [ \Delta x  \; \Delta \xi]^T$ and 
$$ P = \left [ \begin{array}{cc} -\frac{C}{2} & 0 \\ 0 & -\frac{L}{2} \end{array} \right ] ,
$$
the interconnection satisfies the dissipation inequality
 \begin{multline}
    \begin{bmatrix}
      \Delta {\dot{z}} \\
      \Delta {z}
    \end{bmatrix}^{\top}
    \begin{bmatrix}
      0 & P 
      \\
      P & 2 \lambda P 
    \end{bmatrix}
    \begin{bmatrix}
      \Delta {\dot{z}}
      \\
      \Delta {z}
    \end{bmatrix}
    \leq  
    \\
    \frac{1}{2}
  \begin{bmatrix}
    \Delta {i}^{cc} \\ \Delta {v}^{cc}
  \end{bmatrix}^{\top}
  \begin{bmatrix}
    -2 \lambda L & -1
    \\
    -1 & 2(G^{d} - \lambda C)
  \end{bmatrix}
  \begin{bmatrix}
    \Delta {i}^{cc} \\ \Delta {v}^{cc}
  \end{bmatrix}
\end{multline}
The storage has a dominance degree 2 and the supply has a negative signature ${\mathcal I} =-1$. 
When terminated, that is, when $i^{cc} = 0$, the circuit is 2-dominant for  $G^{d} < \lambda C $. 
It is a prototype of negative resistance nonlinear oscillator, such as the circuits studied by
Van der Pol \cite{vanDerPol1926} and Nagumo \cite{nagumo1962}.

The series interconnection in Figure \ref{fig:basic:osc} can be studied in a similar way,
as a neutral interconnection between the nonlinear $RL$ circuit in the previous section and a
lossless capacitor. The circuit is signed dissipative with the same storage and with the supply
$$
\sigma(\Delta_{i}, \Delta v) =  \frac{1}{2}
\begin{bmatrix}
  \Delta i^{vc} \\ \Delta v^{vc}
\end{bmatrix}^{\top}
\begin{bmatrix}
   2(R^{d} -\lambda L) & -1
    \\
    -1 &  -2 \lambda C
  \end{bmatrix}
  \begin{bmatrix}
    \Delta {i}^{vc} \\ \Delta {v}^{vc}
      \end{bmatrix}
    $$

\section{Dissipative interconnections}
\label{section:circuits:connection}

\begin{figure}[htpb]
  \centering
  \includegraphics[trim={0.4cm, 0.25cm, 0.3cm, 0.25cm}, clip]{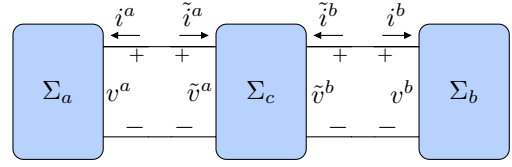}
  \caption{Dissipative interconnection of circuits $\Sigma_{a}$ and $\Sigma_{b}$ through the coupling network $\Sigma_{c}$.}
  \label{fig:dissipative:interconnection}
\end{figure}

We return to question of realizing  dissipative interconnections satisfying (\ref{eq:dissipative:connection}). We illustrate the construction with 
 the \emph{static} coupling network shown in Figure \ref{fig:dissipative:interconnection}.
The interconnection equations are
\begin{align}
  \nonumber
  i^{k} & = - \tilde{i}^{k} + i^{k, cc}, & \tilde{i}^{k} & = - i^{k, vc}
  \\
  v^{k} & = \tilde{v}^{k} + v^{k, vc}, & v^{k} & = v^{k, cc}
  \label{eq:connection:pattern}
\end{align}
where the variables $i^{k, cc}$, $v^{k, cc}$, $i^{k, vc}$ and $v^{k, vc}$, $k \in \{a, b \}$, represent
the range of possible ports available after interconnection.
With this notation, a port is closed or terminated when $i^{k, cc} = 0$ and $v^{k, vc} = 0$, $k \in \{a, b \}$ which 
is the case shown in Figure \ref{fig:dissipative:interconnection}.

The following theorem provides conditions on the coupling network $\Sigma_{c}$ guaranteeing a dissipative interconnection.

\begin{thm}
  \label{thm:dissipative:coupling}
  The interconnection between $\Sigma_{a}$ and $\Sigma_{b}$ is dissipative if and only if the coupling network $\Sigma_{c}$ is 
  signed-passive without any shortage of signed-passivity, i.e., if and only if $\Sigma_{c}$ satisfies,
  \begin{equation}
    0 \leq  
    \begin{bmatrix}
      \Delta \tilde{i}^{a} \\ \Delta \tilde{i}^{b} \\ \Delta \tilde{v}^{a} \\ \Delta \tilde{v}^{b}
    \end{bmatrix}^{\top}
    \begin{bmatrix}
      \tilde{\mathcal{Q}}_{a} & 0 & \mathcal{I}_{a} & 0
      \\
      0 & \tilde{\mathcal{Q}}_{b} & 0 & \mathcal{I}_{b}
      \\
      \mathcal{I}_{a} & 0 & \tilde{\mathcal{R}}_{a} & 0 
      \\
      0 & \mathcal{I}_{b} & 0 & \tilde{\mathcal{R}}_{b}
    \end{bmatrix}
    \begin{bmatrix}
      \Delta \tilde{i}^{a} \\ \Delta \tilde{i}^{b} \\ \Delta \tilde{v}^{a} \\ \Delta \tilde{v}^{b}
    \end{bmatrix}
    \label{eq:coupling:passive}
  \end{equation}
  with $\tilde{\mathcal{Q}}_{k} \leq 0$, $\tilde{\mathcal{R}}_{k} \leq 0$ for all $k \in \{a, b\}$. In addition, the interconnection is
  neutral if and only if,
  \begin{equation}
    0 = \Delta \tilde{i}^{a} \mathcal{I}_{a} \Delta \tilde{v}^{a} + \Delta \tilde{i}^{b} \mathcal{I}_{b} \Delta \tilde{v}^{b}
    \label{eq:coupling:neutral}
  \end{equation}
\end{thm}

\begin{pf}
  Computation of the left-hand side of \eqref{eq:dissipative:connection} under the 
  interconnection pattern \eqref{eq:connection:pattern} lead us to,
\begin{align*}
  & \Delta i^{a} \mathcal{I}_{a} \Delta v^{a} + \Delta i^{b} \mathcal{I}_{b} \Delta v^{b} 
  \\
  & = \sum_{k \in \{a, b \}} \left( -\Delta \tilde{i}^{k} + \Delta i^{k, cc} \right) \mathcal{I}_{k} \Delta v^{k}
  \\
  & = \sum_{k \in \{a, b \}} - \Delta \tilde{i}^{k} \mathcal{I}_{k} \left( \Delta \tilde{v}^{k} + \Delta v^{k, vc} \right) +
  \Delta i^{k, cc} \mathcal{I}_{k} \Delta v^{k, cc}
  \\
  & = \sum_{k \in \{a, b \}}  - \Delta \tilde{i}^{k} \mathcal{I}_{k} \Delta \tilde{v}^{k}
  \\
  & \qquad + \sum_{k \in \{a, b\}} \Delta i^{k, cc} \mathcal{I}_{k} \Delta v^{k, cc} + \Delta i^{k, vc} \mathcal{I}_{k} \Delta v^{k, vc}
  \\
  & \leq \sum_{k \in \{a, b\}} \Delta i^{k, cc} \mathcal{I}_{k} \Delta v^{k, cc} + \Delta i^{k, vc} \mathcal{I}_{k} \Delta v^{k, vc} 
\end{align*}
where we have made use of \eqref{eq:coupling:passive} in the last step.
Hence, the conclusion follows by taking 
\begin{align}
  \nonumber
  i & = [i^{a, cc}, i^{b, cc}, i^{a, vc}, i^{b, vc}]^{\top}
  \\
  v  &= [v^{a, cc}, v^{b, cc}, v^{a, vc}, v^{b, vc}]^{\top} 
  \label{eq:vi:vector}
\end{align}
and $\mathcal{I} = \diag [\mathcal{I}_{a}, \mathcal{I}_{b}, \mathcal{I}_{a}, \mathcal{I}_{b} ]$. $\hfill\qed$
\end{pf}

The addition of the network $\Sigma_{c}$ adds \emph{signed} dissipation to both systems, allowing the following
generalization of  Theorem \ref{thm:dominance:closedLoop}.

\begin{cor}
Let  $\Sigma_{a}$ be a strictly signed-passive circuit with rate $\lambda > 0$ and dominance degree $p$.  
The terminated circuit built from dissipative interconnection of $\Sigma_{a}$  with a resistor ($\Sigma_{b}$) through
a coupling $\Sigma_{c}$ defines a  
  $p$-dominant system with the same rate $\lambda > 0$ provided that
  \begin{equation}
    \sum_{k \in \{a, b \}}
    \begin{bmatrix}
      \Delta i^{k} \\ \Delta v^{k}
    \end{bmatrix}^{\top}
    \begin{bmatrix}
      \mathcal{Q}_{k} + \tilde{\mathcal{Q}}_{k} & 0
      \\
      0 & \mathcal{R}_{k} + \tilde{\mathcal{R}}_{k}
    \end{bmatrix}
    \begin{bmatrix}
      \Delta i^{k} \\ \Delta v^{k}
    \end{bmatrix} \leq 0
    \label{eq:dominance:condition:gral}
  \end{equation}
\end{cor}

\begin{pf}
  The proof is the same as in Theorem \ref{thm:dominance:closedLoop} but considering Theorem \ref{thm:dissipative:coupling} and
  the interconnection pattern \eqref{eq:connection:pattern} instead. $\hfill\qed$
\end{pf}

\begin{figure}[htpb]
  \centering
  \includegraphics[trim={0.4cm, 0cm, 0.3cm, 0.2cm}, clip]{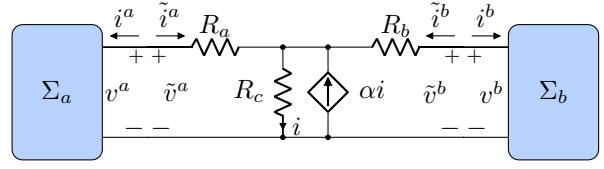}
  \caption{``T'' interconnection of systems $\Sigma_{a}$ and $\Sigma_{b}$ using a current-controlled current source
      for the cases when $\mathcal{I}_{a} = -\mathcal{I}_{b}$.}
  \label{fig:dissipative:T}
\end{figure}

\begin{figure}[htpb]
  \centering
  \includegraphics[trim={0.4cm, 0cm, 0.3cm, 0.2cm}, clip]{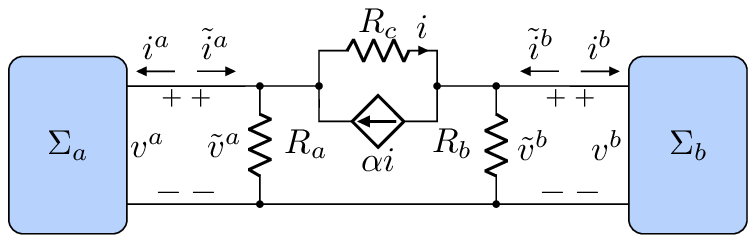}
  \caption{``$\Pi$'' interconnection of systems $\Sigma_{a}$ and $\Sigma_{b}$ using a current-controlled current source
      for the cases when $\mathcal{I}_{a} = -\mathcal{I}_{b}$.}
  \label{fig:dissipative:Pi}
\end{figure}

Figures \ref{fig:dissipative:T}-\ref{fig:dissipative:Pi} illustrate practical realizations of dissipative interconnections
where resistive elements model power losses. 

The ``T'' connection in Figure \ref{fig:dissipative:T} imposes the constraints
\begin{align*}
  i^{a} & = -\tilde{i}^{a}, \quad i^{b} = -\tilde{i}^{b}
  \\
  v^{a} & = \tilde{v}^{a} = R_{a} \tilde{i}^{a} - \frac{R_{c}}{\alpha - 1}(\tilde{i}^{a} + \tilde{i}^{b})
  \\
  v^{b} & = \tilde{v}^{b} = R_{b} \tilde{i}^{b} - \frac{R_{c}}{\alpha - 1}(\tilde{i}^{a} + \tilde{i}^{b})
\end{align*}
where $\alpha > 1$. Without loss of generality we assume that $\mathcal{I}_{a} = -1$ and $\mathcal{I}_{b} = 1$.
It follows from direct computations that the ``T'' bridge satisfies \eqref{eq:coupling:passive} with
\begin{align*}
  \tilde{\mathcal{Q}}_{a} & = R_{a} - \frac{R_{c}}{\alpha - 1}, & \tilde{\mathcal{R}}_{a} & = 0
  \\
  \tilde{\mathcal{Q}}_{b} & = \frac{R_{c}}{\alpha - 1} - R_{b}, & \tilde{\mathcal{R}}_{b} & = 0
\end{align*}

Hence, according to Theorem \ref{thm:dissipative:coupling}, the interconnection of $\Sigma_{a}$ and $\Sigma_{b}$ via the ``T'' bridge
is dissipative for the case $\mathcal{I}_{a} = -1$ and $\mathcal{I}_{b} = 1$ whenever $R_{a} \leq \frac{R_{c}}{\alpha -1} \leq R_{b}$.

The dual version of the ``T'' connection in Figure \ref{fig:dissipative:T} is the ``$\Pi$'' connection
as shown in Figure \ref{fig:dissipative:Pi}.

In this case the connection imposes the relations
\begin{align*}
  v^{a} & = \tilde{v}^{a}, \quad  v^{b} = \tilde{v}^{b}
  \\
  -i^{a} & = \tilde{i}^{a} = \frac{1}{R_{a}} \tilde{v}^{a} - \frac{\alpha - 1}{R_{c}} \left( \tilde{v}^{a} - \tilde{v}^{b} \right)
  \\
  -i^{b} & = \tilde{i}^{b} = \frac{1}{R_{b}} \tilde{v}^{b} + \frac{\alpha - 1}{R_{c}} \left( \tilde{v}^{a} - \tilde{v}^{b} \right)
\end{align*}
where $\alpha > 1$. Hence direct computations show that the ``$\Pi$'' bridge 
also satisfies \eqref{eq:coupling:passive} with 
\begin{align*}
  \tilde{\mathcal{Q}}_{a} & = 0, & \tilde{\mathcal{R}}_{a} & = \frac{1}{R_{a}} - \frac{\alpha - 1}{R_{c}}
  \\
  \tilde{\mathcal{Q}}_{b} & = 0, & \tilde{\mathcal{R}}_{b} & = \frac{\alpha - 1}{R_{c}} - \frac{1}{R_{b}}
\end{align*}

Following again Theorem \ref{thm:dissipative:coupling}, the 
``$\Pi$'' bridge provides an interconnection that is dissipative whenever  
$\frac{1}{R_{a}} \leq \frac{\alpha - 1}{R_{c}} \leq \frac{1}{R_{b}}$.

\begin{figure}[htpb]
  \centering
  \includegraphics[trim={0.4cm, 0cm, 0.3cm, 0.2cm}, clip]{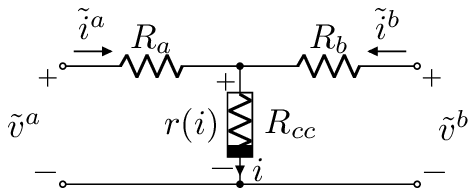}
  \includegraphics[trim={0.4cm, 0cm, 0.3cm, 0.2cm}, clip]{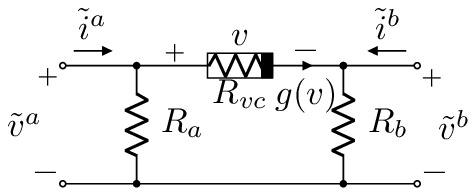}
    \caption{Implementation of dissipative ``T'' and ``$\Pi$'' interconnections via controlled resistors. Both interconnection networks are
    dissipative for systems with opposite supply signature $\mathcal{I}_{a} = -\mathcal{I}_{b}$  in the active range of the controlled resistors.}
  \label{fig:connections:T:Pi:implementation}
\end{figure}

Both dissipative interconnections above can be implemented by using negative resistance devices as shown 
in Figure \ref{fig:connections:T:Pi:implementation}. One should stress that the implementations in Figure \ref{fig:connections:T:Pi:implementation}
only consider the active range of the controlled resistors $R_{vc}$ and $R_{cc}$.

\section{An example}

\begin{figure*}[htpb]
  \centering
  \includegraphics[trim={0.4cm, 0cm, 0.3cm, 0.2cm}, clip]{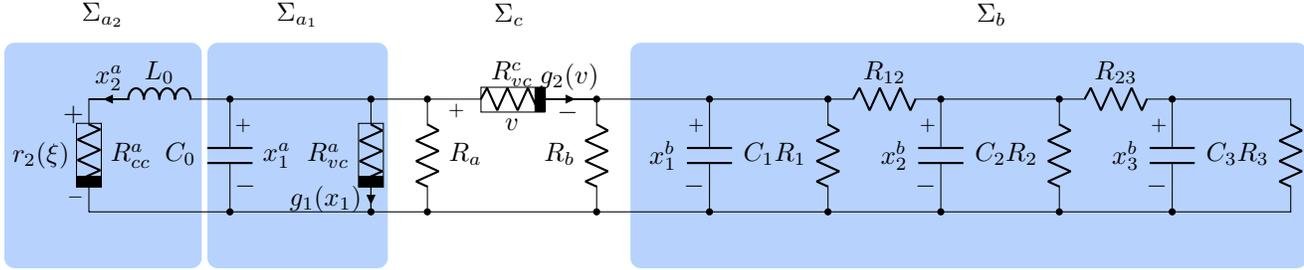}
  \caption{Negative resistance oscillator connected to a passive load through a ``$\Pi$'' dissipative interconnection.}
  \label{fig:fn}
\end{figure*}

We conclude this paper with an analysis of the circuit shown in Figure \ref{fig:fn}. The circuits $\Sigma_{a_{1}}$ and $\Sigma_{a_{2}}$ are the
negative resistance switches analyzed in Section \ref{section:circuits:lossless:lure}. 
From \eqref{eq:cc_sw:ineq}-\eqref{eq:vc_sw:ineq} it becomes clear that their interconnection (denoted as $\Sigma_{a}$) is 
neutral. In addition, Theorem \ref{thm:interconnection} reveals that the resulting circuit is signed-passive with a negative storage
(of dominance degree 2) and a passivity supply with negative signature $-1$, for all 
$\lambda >\max\{ \frac{G^{d}}{C_0}, \frac{R^{d}}{L_{0}} \}$, where $G^{d}$ and $R^{d}$ are the positive slopes of the voltage-current characteristics
of $R_{cc}^{a}$ and $R_{vc}^{a}$ respectively. 

The circuit $\Sigma_{b}$ is a classical linear $RC$ passive load. It has a positive definite storage and is passive, that is signed-passive with
 positive signature supply $+1$, for $\lambda <  \min_{k \in \{1, 2, 3 \} } \left\{ \frac{1}{R_{k} C_{k}} \right\}$.
 
 The two circuits are interconnected through the ``$\Pi$'' bridge discussed in the previous section. This element makes the interconnection 
 of $\Sigma_{a}$ and $\Sigma_{b}$ dissipative. As a consequence, the interconnected circuit is signed-passive. Its storage is the difference
 of two positive definite storages. It has a dominance degree 2. The supply of the interconnected system is a passivity supply with
 positive signature $+1$. The terminated circuit is 2 dominant for any rate $\lambda$ satisfying
 $$\max \left\{ \frac{G^{d}}{C_0}, \frac{R^{d}}{L_{0}} \right \} < \lambda <  \min_{k \in \{1, 2, 3 \} } \left\{ \frac{1}{R_{k} C_{k}} \right\}.$$

\begin{figure}[htpb]
  \centering
  \begin{tikzpicture}
\begin{axis}[
xtick = {0.0, 0.1, 0.2},
xticklabels = {0.0, 0.1, 0.2},
ytick = {-30, -15, 0, 10},
yticklabels = {-30, -15, 0, 10},
x grid style={white!80.0!black},
xlabel={$t [s]$},
xmajorgrids,
xmin=-0.01, xmax=0.21,
y grid style={white!80.0!black},
ylabel={$x_{3}^{b} [mV]$},
ymajorgrids,
ymin=-30, ymax=10,
height=3.5cm, width=0.45\textwidth,
]
\addplot [thick, black, forget plot] table [x=t, y=Y2, col sep=comma]{nagumo-passive2.csv};
\end{axis}
\end{tikzpicture}
\caption{Time trajectory of the voltage across the capacitor $C_{3}$ of the circuit in Figure \ref{fig:fn}.}
\label{fig:fn:trajectories:dissipative}
\end{figure}
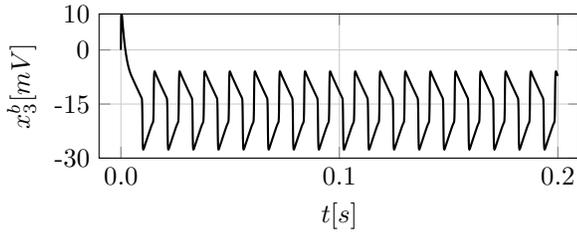

The simulation in Figure \ref{fig:fn:trajectories:dissipative} is for the set of parameters $L_{0} = 50 mH$, 
$C_{0} = 10 \mu F$, $C_{1} = C_{2} = C_{3} = 0.1 \mu F$, $R_{1} = R_{2} = R_{3} = R_{12} = R_{23} = 1 \Omega$,
$R_{a} = 20 \Omega$, and $R_{b} = 10 \Omega$.
The active resistors $R_{vc}^{a}$, $R_{cc}^{a}$ and $R_{vc}^{c}$ have voltage-current characteristics given by
\begin{align*}
  g_{1}(x_{1}) &= 
  \begin{cases}
    0.1 x_{1} & x_{1} < 2 V
    \\
    -0.1 x_{1} + 0.4 & 2 V \leq x_{1} \leq 3 V
    \\
    0.1 x_{1} - 0.2 & 3V < x_{1}
  \end{cases}
  \\
  r_{2}(x_{2}) &= 
  \begin{cases}
    10 x_{2} + 5 & x_{2} < -0.2 A
    \\
    -10 x_{2} + 1 & -0.2A \leq x_{2} \leq -0.1 A
    \\
    10 x_{2} + 3 & -0.1A < x_{2}
  \end{cases} 
  \\
  g_{2}(v) & = 
  \begin{cases}
    0.1375 v + 0.9625 & v < -5 V
    \\
    -0.055 v & -5 V \leq v \leq 5 V
    \\
    0.1375 v - 0.9625 & 5 V \leq v
  \end{cases}
\end{align*}
Note that the active resistor $R_{vc}^{c}$ has an active region with negative slope of $-0.055$ and satisfies 
$\frac{1}{R_{a}} \leq 0.055 \leq \frac{1}{R_{b}}$, thus providing a dissipative coupling locally. 
Also, with these set of  parameters the circuit has a unique unstable equilibrium. The simulated behavior is bounded 
and entirely in the active range of the controlled resistors. 
By 2-dominance of the circuit, the trajectory must converge to a limit cycle.

\linespread{0.98}

\bibliography{Biblio}
\bibliographystyle{plain}

\end{document}